\documentclass[12pt,letterpaper]{article}

\usepackage{amsmath,amssymb,array,calc,rotating,epsfig,psfrag, amscd}
\usepackage{color}
\usepackage{hyperref}
\usepackage[left=2cm,top=2cm,right=3cm,nohead]{geometry}

\newcommand{\bea}{\begin{eqnarray}}
\newcommand{\eea}{\end{eqnarray}}
\newcommand{\be}{\begin{equation}}
\newcommand{\ee}{\end{equation}}
\newcommand{\barr}{\begin{array}}
\newcommand{\earr}{\end{array}}

\newcommand{\bpm}{\begin{pmatrix}}
\newcommand{\epm}{\end{pmatrix}}
 \newcommand{\bitem}{\begin{itemize}}
 \newcommand{\eitem}{\end{itemize}}

\definecolor{cardinal}{rgb}{0.6,0,0}
\definecolor{darkgreen}{rgb}{0,0.5,0}
\definecolor{golden}{rgb}{0.92, 0.7, 0}
\definecolor{midnight}{rgb}{0, 0, 0.5}
\definecolor{darkblue}{rgb}{0.2, 0, 0.8}

\newcommand{\beq}{\begin{equation}\begin{aligned}}
\newcommand{\eeq}{\end{aligned}\end{equation}}

\begin{document}

\vspace{0.5cm}
\begin{center}

{\Large \bf Remark on the Baryonic Branch\\ of the Warped Deformed Conifold}\\
\vskip 3mm

 \vskip1.5cm 
Gregory Giecold\footnote{\href{mailto:giecold@insti.physics.sunysb.edu}{{\tt giecold@insti.physics.sunysb.edu}}} 
 \vskip0.7cm
\textit{C.N.~Yang Institute for Theoretical Physics,\\
State University of New York,\\
Stony Brook, NY 11794--3840\\ U.S.A.}\\
\end{center}
\vskip1.5cm
\begin{abstract}

It has recently been suggested that a superpotential for the baryonic branch of the Klebanov--Strassler field theory (KS) should exist for a sector of the dual supergravity fields. 
In this note we would like to argue that if extended to cover the whole set of supergravity fields a would--be superpotential does not have a perturbative expansion around the known KS superpotential. Since the family of supergravity duals to the baryonic branch is an expansion around the warped deformed conifold, our argument most likely indicates that there is no such superpotential, and hints that some one--parameter supersymmetric solutions do not arise from BPS flow equations. 
\end{abstract} 

\newpage

\section{Introduction}

The warped deformed conifold~\cite{arXiv:hep-th/0007191} (also known as the Klebanov--Strassler solution), the solution corresponding to D3--branes on the resolved conifold~\cite{hep-th/0010088} and the Chamseddine--Volkov/Maldacena--Nu\~nez solution~\cite{hep-th/9707176, hep-th/9711181, hep-th/0008001} (CVMN), all three arise from a common reduction of IIB supergravity. Indeed, the aforementioned solutions -- realizing the holographic dual to confinement and chiral symmetry breaking via the deformation of the conifold, or whose field theory dual flows in the infrared to four--dimensional $\mathcal{N}=1$ sYM --  are particular solutions to the one--dimensional Lagrangian obtained by reduction of the Papadopoulos--Tseytlin Ansatz (PT)~\cite{hep-th/0012034} for IIB supergravity. This Ansatz underlies the family of supersymmetric, regular solutions found by Butti, Gra\~na, Minasian, Petrini and Zaffaroni~\cite{hep-th/0412187} describing the baryonic branch of the Klebanov--Strassler field theory dual~\cite{hep-th/0101013, hep-th/0511254}\footnote{See~\cite{hep-th/0405282, hep-th/0409186} for work that anticipated this result and~\cite{arXiv:0803.1315} for a short review with physical motivations and subsequent references.} 

As stressed in~\cite{hep-th/0012034, arXiv:1111.6567, arXiv:1008.0983}, it would be natural to try and find a superpotential for the generic solutions interpolating between the Klebanov--Strassler and the CVMN solutions, especially in view of the extensive use made of superpotentials\footnote{Following the approach initiated by Borokhov and Gubser~\cite{hep-th/0206098}.} in the analysis of the space of linearized perturbations around backgrounds with charge dissolved in flux, the identification of candidate duals to metastable supersymmetry--breaking states and the issue of their infrared singularities~\cite{hep-th/0309011, arXiv:0912.3519, arXiv:1106.6165, arXiv:1102.2403, arXiv:1102.1734, arXiv:1011.2626, arXiv:1011.2195, arXiv:1108.1789, arXiv:1110.2513}. See also~\cite{arXiv:1106.0002, arXiv:1111.1727, arXiv:1111.1731} and~\cite{arXiv:1105.4879, arXiv:1111.2605} for related investigations.

The aim of this note is to explain why finding such an interpolating superpotential likely cannot be achieved. In turn, this would imply that there exist supersymmetric solutions that depend on one variable and yet fail to admit a superpotential, which is rather unexpected in view of the standard lore. More specifically, the family of supersymmetric solutions~\cite{hep-th/0412187} interpolating between CVMN and the warped deformed conifold are determined by first--order equations; our result implies that those first--order equations are not flow equations derived from a superpotential.

It is important to note that the results of the present paper are not in contradiction with a recent indirect argument pointing to the existence of a superpotential for the NS sector of the supergravity dual to the baryonic branch of KS~\cite{arXiv:1111.6567}. The authors of~\cite{arXiv:1111.6567} have indeed rediscovered the superpotential for a particular NS--sector truncation on the conifold; this superpotential was actually first derived in~\cite{HoyosBadajoz:2008fw}. 
On the other hand, one can generate the baryonic branch for the Klebanov--Strassler solution from Type I supergravity, applying the TST transformation used by Maldacena and Martelli to this purpose~\cite{arXiv:0906.0591}.
The Type I Ansatz used by Maldacena and Martelli satisfies the BPS flow equations derived from the superpotential for the NS--sector truncation of~\cite{arXiv:1111.6567}.
The proposal of~\cite{arXiv:1111.6567} is that in principle a superpotential for the NS fields of the baryonic branch could then be obtained by applying the TST transformation of Maldacena and Martelli. This amounts to a field redefinition and even though there was some confusion after the release of~\cite{arXiv:1111.6567} shortly before the work at hand, it is actually not claimed in~\cite{arXiv:1111.6567} that type I fields could somehow be morphed into the full set of supergravity fields describing the baryonic branch of KS. 

\section{The Papadopoulos--Tseytlin Ansatz}

The Klebanov--Strassler background~\cite{arXiv:hep-th/0007191} exhibits an $\text{SU}(2) \times \text{SU}(2) \times \mathbb{Z}_2$ symmetry. The $\mathbb{Z}_2$ symmetry interchanges the angular coordinates $\left( \theta_1, \phi_1 \right)$ and $\left( \theta_2, \phi_2 \right)$ parameterizing the two $S^2$'s from the tranverse topology. On the gauge theory side, this symmetry amounts to exchanging the fundamental and anti--fundamental representations of the $\text{SU}(N+M) \times \text{SU}(M)$ gauge groups. Outstanding surveys of this background can be found in~\cite{hep-th/0108101, hep-th/0205100}. 

We are interested in deformations of the warped deformed conifold that break this $\mathbb{Z}_2$ symmetry (whereby the two copies of $\text{SU}(2)$ are inequivalent). As such, we are looking for a family of non--supersymmetric solutions with $\text{SU}(2) \times \tilde{\text{SU}}(2)$ symmetry which are continuously connected to the KS solution. 
 
The most general Ansatz consistent with those symmetries was put forward by Papadopoulos and Tseytlin (PT)~\cite{hep-th/0012034}. The metric is written as
  \beq
  \label{PTmetric}
 ds_{10}^2= \, e^{2A}\, ds_{1,3}^2 + e^{-6p-x}\, d\tau^2 + e^{x+g}\, (e_1^2+e_2^2) + e^{x-g}\, (\tilde{\epsilon}_1^2+\tilde{\epsilon}_2^2) + e^{-6p-x}\, \tilde{\epsilon}_3^2 \ ,
 \eeq
where all the stretching and warping functions depend only on the bulk radial variable $\tau$ and we found it more convenient to work with the conventions of~\cite{hep-th/0412187} instead of those originally used by Papadopoulos and Tseytlin. In particular, we opt for a string--frame metric. Note that the two $S^2$'s become equivalent when $a^2 = 1 - e^{2 g}$. This would reduce the number of functions in the Ansatz by one, $a = \tanh(y)$, $e^{-g} = \cosh(y)$.

The fluxes and axio--dilaton of the PT Ansatz are
 \begin{align}
 \label{PTfluxes}
 H_3 = & \, h_2(\tau)\, \tilde{\epsilon}_3 \wedge \left( \epsilon_1 \wedge e_1 + \epsilon_2 \wedge e_2 \right) + d\tau \wedge \Big[ h_1'(\tau)\, \left( \epsilon_1 \wedge \epsilon_2 + e_1 \wedge e_2 \right)  \nonumber\\ & + \chi'(\tau)\, \left(- \epsilon_1 \wedge \epsilon_2 + e_1 \wedge e_2 \right) + h_2'(\tau)\, \left( \epsilon_1 \wedge e_2 - \epsilon_2 \wedge e_1 \right) \Big]  \ ,
 \end{align}
 \begin{align}
 F_3 = & \, P\, \Big[ \tilde{\epsilon}_3 \wedge \Big[ \epsilon_1 \wedge \epsilon_2 + e_1 \wedge e_2 - b(\tau)\, \left( \epsilon_1 \wedge e_2 - \epsilon_2 \wedge e_1 \right) \Big] \nonumber\\ & + b'(\tau)\, d\tau \wedge\left( \epsilon_1 \wedge e_1 + \epsilon_2 \wedge e_2 \right) \Big] \ ,
 \end{align}
 \begin{align}
 F_5 = \, {\cal F}_5 + * {\cal F}_5 \ , \ \ \ {\cal F}_5= \Big[ Q + 2\, P\, \big( h_1(\tau) + b(\tau)\, h_2(\tau) \big)\Big] \, e_1 \wedge e_2 \wedge \tilde{\epsilon}_1 \wedge \tilde{\epsilon}_2 \wedge \tilde{\epsilon}_3  \ ,\nonumber\\
 \end{align}
 \beq
 \Phi = \Phi(\tau) \ ,\ \ \ C_0 =0 \, , \nonumber\\
 \eeq
with $Q,P$ being related to the number of ordinary and fractional branes, respectively. A prime denotes a derivative with respect to $\tau$. The function $\chi$ is an additional component of the NS three--form which arises from breaking the $\mathbb{Z}_2$ symmetry of the warped deformed conifold. It is commonly eliminated via its algebraic equation of motion.

The IIB supergravity action is then reduced to a one--dimensional effective action that gives the equations of motion for the fields entering the Papadopoulous--Tseytlin Ansatz. This takes the following form:
\beq\label{Lag1}
\mathcal{S}_1 = \, \int d\tau \left( - \frac{1}{2}\, G_{ab}\, \phi^{\prime a}\, \phi^{\prime b} - V(\phi) \right) \ .
\eeq
The set of scalar functions $\phi^a$, $a=1,...,10$ appearing in the above Ansatz will from now on be referred to in the following order:
\beq
\label{phidef}
 \phi^a= \big( a, g, x, p, A, \Phi, b, h_1, h_2, \chi \big) \ .
 \eeq
The field--space metric is found to be
\begin{align}
\label{fieldmetric}
- \frac{1}{2}\, G_{ab}\, \phi^{\prime a}\, \phi^{\prime b} \,= & \,  e^{4 A + 2 x - 2\Phi }\, \Big[ - \frac{1}{4}\, e^{-2 g}\, a'^2 - \frac{1}{4}\, g'^2 + \frac{1}{4}\, x'^2 + 3\, A'^2 + \Phi'^2 + 3\, A'\, x' - 4\, A'\, \Phi ' \nonumber\\ & - 6\, A'\, p' - \frac{3}{2}\, x'\, \Phi ' - 3\, x'\, p' + 3\, p' \, \Phi ' \Big] - \frac{1}{8}\, e^{4 A}\, \Big[ e^{- 2 \Phi }\, \Big[ e^{2 g}\, ( h_1' - \chi ')^2 \nonumber\\ & + e^{-2 g}\, \left( (1+a^2 )\, h_1' + 2\, a\, h_2' + (1-a^2)\, \chi ' \right)^2 + 2\, \left( a\, h_1' + h_2' - a\, \chi ' \right)^2 \Big]+ 2\, P^2\, b'^2 \Big] \, ,\nonumber\\
\end{align}
while the potential is given by
\beq\label{Vpot}
V(\phi) = \, V_{gr}(\phi) + V_{mat}(\phi) \ ,
\eeq
where
\begin{align}\label{Vgr}
V_{gr}(\phi) = & \, -\frac{1}{2}\, e^{4 A - 6 p - 2 \Phi }\, \left(e^g+ (1 + a^2 )\, e^{- g} \right) \nonumber\\
& + \frac{1}{4}\, a^2\, e^{4 A - 2 g + 2 x - 2 \Phi } \nonumber\\ & + \frac{1}{8}\, e^{4 A - 12 p - 2 x - 2 \Phi }\, \left(e^{2 g}+ (a^2 - 1)^2\, e^{-2 g} + 2\, a^2 \right)
\end{align}
and
\begin{align}\label{Vmat}
V_{mat} = & \, \frac{1}{8}\, \Big[ 2\, e^{4 A - 2 \Phi}\, h_2^2 + P^2\, e^{4 A}\, \left( e^{2g} + e^{-2g}\, (a^2-2\, a\, b +1)^2+2\, (a-b)^2 \right)\nonumber\\ & + e^{4 A - 2 x}\, \big( Q+ 2\, P\, (h_1 + h_2\, b) \big)^2 \Big] \ .
\end{align}

There is in addition the ``zero--energy condition'' coming from the $R_{\tau \tau}$ component of the Einstein equations, which results in the constraint $\frac{1}{2}\, G_{ab}\, \phi^{\prime a}\, \phi^{\prime b} = V(\phi)$.

\section{Candidate superpotential}

By definition\footnote{When the warp factor is counted out of the field space metric, this relation is written as $V = \frac{1}{8}\, G^{ab}\, \frac{\partial W}{\partial \phi^a}\, \frac{\partial W}{\partial \phi^b} - \alpha W^2$ for some particular real number $\alpha$.}, a superpotential is related to $V$~\eqref{Vpot} through
\beq\label{Superpotential def}
V = \, \frac{1}{8}\, G^{ab}\, \frac{\partial W}{\partial \phi^a}\, \frac{\partial W}{\partial \phi^b} \ .
\eeq   
Second--order equations of motion and the ``zero--energy'' condition then follow from the system of first--order flow equations
\beq\label{flow eq}
\phi^{\prime a} = \, \frac{1}{2} G^{ab} \frac{\partial W}{\partial \phi^b} \, .
\eeq
Whether or not solutions to these first--order equations are actually BPS is a subtle issue, depending on the superpotential being a genuine superpotential or a fake one. See~\cite{arXiv:1111.6567} for a nice discussion and pointers to a vast literature.

Our purpose is now to try and identify a superpotential for the Papadopoulos--Tseytlin Ansatz. Let us outline how we proceed. It would clearly be hopeless and overkill to give a stab at solving an eikonal equation for an unknown function $W(\phi^a)$ with an entirely undetermined dependence on the fields $\{ \phi^a \}, \ a = 1, ..., 10$. Rather, the first stage to ease the task is to note that many of the fields $\phi^a$ appear only as exponentials affected with some specific weights. 

For instance, the warp factor $A$ from~\eqref{PTmetric} comes into sight only as a global $e^{4 A}$ everywhere in the potential~\eqref{Vpot} of the Papadopoulos--Tseytlin Ansatz. Similarly, the inverse metric involves an $e^{- 4 A}$ in all its entries. Combining these observations with relation~\eqref{Superpotential def} guarantees that the field $A$ is bound to appear in the superpotential as an overall $e^{4 A}$. 
Following the same reasoning for the fields $p$, $x$ and $\Phi$, all of which show up only as exponentials of definite weights in the field--space metric and Papadopoulos--Tseytlin potential, we are led to consider the following contender for a superpotential:

\begin{align}\label{Wcan}
W_{candidate} = & \, e^{4A-2\Phi}\, \left[ e^{-6p}\, \sqrt{1+\frac{1}{4}\, e^{-2g}\, \left( 1 - a^2 - e^{2g} \right)^2} + e^{2x}\, \lambda[a,g] \right] \nonumber\\ & + \frac{1}{2}\, e^{4A-\Phi}\, \zeta[a,g]\, \Big[ Q + 2\, P\, \big( h_1 + b\, h_2 \big) \Big] \, .
\end{align}

In the forthcoming discussion we explain in some more detail why a would--be superpotential for the baryonic branch of the warped deformed conifold must be of this form. It will be convenient to refer to the different pieces of $W_{candidate}$ as follows
\beq
W_{candidate} = \, W^{(1)} + W^{(2)} + W^{(3)} \, ,
\eeq
where
\begin{align}
& W^{(1)} = \, e^{4A-6p-2\Phi}\, \sqrt{1+\frac{1}{4}\, e^{-2g}\, \left( 1 - a^2 - e^{2g} \right)^2} \, , \nonumber\\
& W^{(2)} = \, e^{4A+2x-2\Phi}\, \lambda[a,g] \, , \nonumber\\
& W^{(3)} = \, \frac{1}{2}\, e^{4A-\Phi}\, \zeta[a,g]\, \Big[ Q + 2\, P\, \big( h_1 + b\, h_2 \big) \Big] \, .
\end{align}

In view of the expression derived from~\eqref{fieldmetric} for the inverse field--space metric $G^{ab}$, it is clear that only $G^{ab}\, \frac{\partial W}{\partial \phi^a}\, \frac{\partial W}{\partial \phi^{b}}$ with derivatives of $W_{candidate}$ acting solely on $W^{(1)}$ have a chance to reproduce the term of $V$~\eqref{Vpot} that is proportional to $e^{4A-12p-2x-2\Phi}$, see the third line of~\eqref{Vgr}. 
Similarly, derivatives acting on $W^{(2)}$ are the only ones that have any chance to give rise to the term $\frac{1}{4}\, a^2\, e^{4 A - 2 g + 2 x - 2 \Phi}$ on the second line of~\eqref{Vgr}, once again from consideration of the weights in $A$, $x$ and $\Phi$.

The whole term on the second line of~\eqref{Wcan}, i.e. $W^{(3)}$, should give rise to all of $V_{mat}$. Indeed, there is no other possibility. For instance, cross--terms of the type $G^{ab}\, \frac{\partial W^{(i)}}{\partial \phi^{a}}\, \frac{\partial W^{(j)}}{\partial \phi^{b}}$, $(i \neq j)$ cannot possibly yield the correct weights in $A$, $x$ and $\Phi$ found in $V_{mat}$. Explicitly, we see that acting with a field derivative on $W^{(3)}$ results in an overall factor of $e^{4A-\Phi}$. Acting on $W^{(1)}$ or $W^{(2)}$ generates an overall $e^{4A-2\Phi-6p}$ or an overall $e^{4A-2\Phi+2x}$. The possible $G^{ab}$'s linking those two sets of derivatives give either $e^{-4A}$ or something proportional to $e^{-4A-2x+2\Phi}$. We thus see that no cross--term can possibly reproduce {\it any} of the weights appearing in $V_{mat}$, namely $e^{4A}$, $e^{4A-2\Phi}$ or $e^{4A-2x}$.

Proceeding further in this stepwise way, it can be verified that $W^{(1)}$ correctly reproduces the last term in the potential~\eqref{Vgr}, i.e.
\begin{align}\label{W3 eik}
\frac{1}{8}\, G^{ab}\, \frac{\partial W^{(1)}}{\partial \phi^a}\, \frac{\partial W^{(1)}}{\partial \phi^b} = \, \frac{1}{8}\, e^{4 A - 12 p - 2 x - 2 \Phi }\, \left(e^{2 g}+ (a^2 - 1)^2\, e^{-2 g} + 2\, a^2 \right) \ .
\end{align}

A hint that led us in the first place to this expression for $W^{(1)}$ boils down to it being proportional to the known superpotential for the Maldacena--Nu\~nez solution (as first identified in Section 5.2 of~\cite{hep-th/0012034}). Furthermore, we will see in a short while that on the $\mathbb{Z}_2$--symmetric point of the baryonic branch it reduces --- as it should --- to one of the pieces from the known expression for the Klebanov--Strassler superpotential.

Now that such considerations on the weights of the exponentials of $A$, $p$, $x$ and $\Phi$ have cleared quite a lot the allowed structure of a candidate superpotential, one should next determine $\lambda[a,g]$ and $\zeta[a,g]$ entering~\eqref{Wcan} by requiring that the defining equation for a superpotential~\eqref{Superpotential def} be obeyed, with the potential $V$ given in~\eqref{Vgr} and~\eqref{Vmat}. 

However, in view of an argument we have already appealed to --- namely that no cross-term can possibly be involved --- one can first apply another preemptive simplification before embarking on this task. Indeed, we notice that setting $\zeta[a,g] \equiv 1$ is such that the second line of $W_{candidate}$ all by itself correctly accounts for the whole of $V_{mat}$. 

Accordingly, we are now ready to insert
\begin{align}\label{Wcan2}
W_{candidate} = & \, e^{4A-2\Phi}\, \left[ e^{-6p}\, \sqrt{1+\frac{1}{4}\, e^{-2g}\, \left( 1 - a^2 - e^{2g} \right)^2} + e^{2x}\, \lambda[a,g] \right] \nonumber\\ & + \frac{1}{2}\, e^{4A-\Phi}\, \Big[ Q + 2\, P\, \big( h_1 + b\, h_2 \big) \Big] \, 
\end{align}
into the defining relation~\eqref{Superpotential def}. As it turns out, one ends up with {\it two} partial differential equations to solve, including an eikonal equation for the unknown function $\lambda[a,g]$:
\beq\label{eikonal f}
e^{2 g}\, \left(\frac{\partial \lambda}{\partial a}\right)^2 + \left( \frac{\partial \lambda}{\partial g} \right)^2 = \, a^2\, e^{- 2 g} \ .
\eeq
It is not so difficult to guess the solution to this eikonal equation\footnote{A two--dimensional eikonal equation is a first--order, nonlinear partial differential equation of the form $u_x^2 + u_y^2 = n(x,y)^2$. The surfaces $u(x,y) = c$ are the wavefronts, $n(x,y)$ corresponds to the ``refraction of the medium''.}. Its expression is quite neat:
\beq\label{lambdeikonal}
\lambda[a,g] =\, \sqrt{1 + a^2\, e^{-2 g}} \, .
\eeq
Alternatively, this equation can be solved using the method of characteristics~\cite{PDEs, PDEs2}, taking the Klebanov--Strassler solution as the parameterized initial curve $\Gamma(s): (a(0,s), g(0,s), f(0,s))$. By the uniqueness theorem for solutions to p.d.e.'s of the eikonal type, our expression~\eqref{lambdeikonal} for $\lambda[a,g]$ is the only acceptable solution to~\eqref{eikonal f} that goes through the $\mathbb{Z}_2$--symmetric point of the baryonic branch (i.e.~the Klebanov--Strassler solution~\cite{arXiv:hep-th/0007191}). 

Note that if the fields $a$ and $g$ from the Papadopoulos--Tseytlin Ansatz are constrained by the $\mathbb{Z}_2$ symmetry relation $a^2 = 1 - e^{2 g}$ (so that now $a = \tanh(y)$ and $e^{-g} = \cosh(y)$) then $W^{(2)}$ reduces to $W^{(2)} \rightarrow \cosh(y)\, e^{4 A + 2 x - 2 \Phi}$. Similarly, $W^{(1)}$ becomes $e^{4 A - 6 p - 2 \Phi - g}$. This way, $W_{candidate}$ indeed reduces to the known superpotential for the warped deformed conifold, first found in~\cite{hep-th/0012034}\footnote{After taking into account that in~\cite{hep-th/0012034}, the superpotential is written in Einstein frame and with a different choice of the warp factor multiplying the Minkowski part of the 10d Ansatz metric.}.\\ 

It would thus naively appear that we have obtained a strong candidate for the superpotential for the baryonic branch of the Klebanov--Strassler solution. We have seen how each of its three distinctive pieces correctly reproduce separate terms in the PT potential and how, on the $\mathbb{Z}_2$--symmetric point of the baryonic branch, they yield the known expression for the KS superpotential. 

But this is not the end of the story and it turns out that $W_{candidate}$, namely
\begin{align}\label{full W}
W_{candidate} = \, & e^{4 A + 2 x - 2 \Phi }\, \sqrt{1 + a^2\, e^{- 2 g}} + \frac{1}{2}\, e^{4 A - 6 p - 2 \Phi - g}\, \sqrt{a^4+2\, a^2\left(-1 + e^{2 g}\right)+\left(1 + e^{2 g}\right)^2} \nonumber\\ & + \frac{1}{2}\, \left[ Q + 2 P (h_1 + b\, h_2) \right]\, e^{4 A - \Phi} \, ,
\end{align}
unfortunately fails to satisfy the defining relation $V =\, \frac{1}{8}\, G^{ab}\, \frac{\partial W}{\partial \phi^a}\, \frac{\partial W}{\partial \phi^b}$.
 
Indeed, the partial differential equation~\eqref{eikonal f} that we have solved for $\lambda[a,g]$ is not the only one that is required for the defining relation~\eqref{Superpotential def} to hold. One must also ensure that $\lambda[a,g]$ obeys
\begin{align}\label{other pde}
& 8\, \left[ e^{2g} + \frac{1}{4}\, \left( 1 - a^2 - e^{2g} \right)^2 \right]\, \lambda[a,g] + \left(1-a^2-e^{2 g}\right)\, \left[ \left(1-a^2+e^{2 g}\right)\, \frac{\partial \lambda[a,g]}{\partial g} + 2\, a\, e^{2 g}\, \frac{\partial \lambda[a,g]}{\partial a} \right] \nonumber\\ 
&\, \, \, \, \, \, \, \, \, \, \, \, \, \, \, \, \, \, \, \, \, \, \, \, \, \, \, \, \, \, \, \, \, \, \, \, \, \, \, \, \, \, \, \, \, \, \, \, \, \, \, \, \, \, \, \, \, \, \, \, \, \, \, \, \, \, \, \, \, \, \, \, \, \, \, \, \, \, \, \, \, \, \, \, \, \, \, \, \, \, \, \, \, \, \, \, \, \, \, \, \, \, \, \, \, \, \, \, \, \, \, \, \, \, \, \, \, \, \, \, \, \, \, \, \, \, \, \, \, \, \, \overset{?}{=} \nonumber\\ & \, \, \, \, \, \, \, \,  \, \, \, \, \, \, \, \, \, \, \, \, \, \, \, \, \, \, \, \, \, \, \, \, \, \, \, \, \, \, \, \, \, \, \, \, \, \, \, \, \, \, \, \, \, \, 2\, e^{g}\, \left(1+a^2+e^{2 g}\right)\, \sqrt{2\, \left(1+a^2\right) + \left(-1+a^2\right)^2\, e^{-2 g}+e^{2 g}}  \, . 
\end{align}
The only acceptable solution to the eikonal equation~\eqref{eikonal f} --- that is, the one from equation~\eqref{lambdeikonal} --- fails to satisfy equation~\eqref{other pde}, the other of the two constraints for a superpotential to exist for the baryonic branch, apart from~\eqref{eikonal f} which we successfully solved.

This obstruction stems from the impossibility for the mixing of the derivatives of $W^{(1)}$ and of $W^{(2)}$ to correctly reproduce no more than the first term in the ``metric'' part of the PT potential, the one appearing on the first line of the r.h.s. to equation~\eqref{Vgr}. 

As we have seen, this conclusion is backed by actually solving an eikonal equation for our candidate superpotential\footnote{After deducing its admissible form by some previous consideration explained at length in the bulk of Section 3.}. As a cross--check it should be mentioned that we have also separately verified that the obstruction to getting a superpotential fails at fourth order in a series expansion of the fields from the PT Ansatz around the Klebanov--Strassler solution.

The end--result to the approach exposed in this note --- the lack of a superpotential for the baryonic branch of the warped deformed conifold --- might seem unexpected, especially in view of some unpublished results~\cite{unpub} establishing the existence of a superpotential for a higher--dimensional analogue of the Papadopoulos--Tseytlin Ansatz, encompassing the so--called warped Stenzel background\footnote{For more information on this eleven--dimensional supergravity solution see, e.g.,~\cite{Ceresole:1999zg, Martelli:2009ga, Klebanov:2010qs} and~\cite{arXiv:1011.2195}.}.

Our result does not rule out however that there might be a superpotential for {\it parts of} the baryonic branch, away from the $\mathbb{Z}_2$--symmetric point of the family (the Klebanov--Strassler solution). Indeed, crucial to our argument and to solving the eikonal equation~\eqref{eikonal f} is an initial condition for this p.d.e. While a solution to an Hamilton--Jacobi equation always exist locally\footnote{For more details and a list of references, see~\cite{Chemissany:2010zp} in the context of black hole physics.}, there is no general theorem ensuring its global existence. Yet, we had little choice but to take some of the known expressions for the Klebanov--Strassler solution as our initial conditions, given that the Klebanov--Strassler solution is the only solution among the family of supergravity duals to the baryonic branch for which a superpotential is explicitly known. Our result seems to rule out the existence of a superpotential on a field--space patch centered around KS beyond fourth--order in a series expansion of the supergravity fields around the Klebanov--Strasser solution.

\vskip 0.7cm
 \noindent {\bf Acknowledgements}:\\
 \noindent I am grateful to Iosif Bena for helpful comments on a preliminary version of this note. Following its release on the arXiv, I have benefited from discussions with Anatoly Dymarsky and Thomas van Riet. This work was initiated at IPhT, CEA/Saclay, where this result has been discussed with Iosif Bena, Mariana Gra\~na, Nick Halmagyi, Stefano Massai and Francesco Orsi. I have benefitted from generous support of a Contrat de Formation par la Recherche and an ERC Starting Independent Researcher Grant 240210 -- String--QCD--BH. Financial support by the Research Foundation, Stony Brook University is appreciated.  




\begin{thebibliography}{99}
  
\bibitem{arXiv:hep-th/0007191} 
  I.~R.~Klebanov and M.~J.~Strassler,
  ``Supergravity and a confining gauge theory: Duality cascades and chi SB resolution of naked singularities,''
  JHEP\ {\bf 0008}, 052  (2000)
  \href{http://arxiv.org/abs/hep-th/0007191}{[hep-th/0007191]}.
  
\bibitem{hep-th/0010088} 
  L.~A.~Pando Zayas and A.~A.~Tseytlin,
  ``3-branes on resolved conifold,''
  JHEP\ {\bf 0011}, 028  (2000)
  \href{http://arxiv.org/abs/hep-th/0010088}{[hep-th/0010088]}.
  
\bibitem{hep-th/9707176} 
  A.~H.~Chamseddine and M.~S.~Volkov,
  ``NonAbelian BPS monopoles in N=4 gauged supergravity,''
  Phys.\ Rev.\ Lett.\ \ {\bf 79}, 3343  (1997)
  \href{http://arxiv.org/abs/hep-th/9707176}{[hep-th/9707176]}.
  
\bibitem{hep-th/9711181} 
  A.~H.~Chamseddine and M.~S.~Volkov,
  ``NonAbelian solitons in N=4 gauged supergravity and leading order string theory,''
  Phys.\ Rev.\ D\ {\bf 57}, 6242  (1998)
  \href{http://arxiv.org/abs/hep-th/9711181}{[hep-th/9711181]}.
  
\bibitem{hep-th/0008001} 
  J.~M.~Maldacena and C.~Nunez,
  ``Towards the large N limit of pure N=1 superYang-Mills,''
  Phys.\ Rev.\ Lett.\ \ {\bf 86}, 588  (2001)
  \href{http://arxiv.org/abs/hep-th/0008001}{[hep-th/0008001]}.
  
\bibitem{hep-th/0012034} 
  G.~Papadopoulos and A.~A.~Tseytlin,
  ``Complex geometry of conifolds and five-brane wrapped on two sphere,''
  Class.\ Quant.\ Grav.\ \ {\bf 18}, 1333  (2001)
  \href{http://arxiv.org/abs/hep-th/0012034}{[hep-th/0012034]}.
  
\bibitem{hep-th/0412187} 
  A.~Butti, M.~Grana, R.~Minasian, M.~Petrini and A.~Zaffaroni,
  ``The Baryonic branch of Klebanov-Strassler solution: A supersymmetric family of SU(3) structure backgrounds,''
  JHEP\ {\bf 0503}, 069  (2005)
  \href{http://arxiv.org/abs/hep-th/0412187}{[hep-th/0412187]}.
  
\bibitem{hep-th/0101013} 
  O.~Aharony,
  ``A Note on the holographic interpretation of string theory backgrounds with varying flux,''
  JHEP\ {\bf 0103}, 012  (2001)
  \href{http://arxiv.org/abs/hep-th/0101013}{[hep-th/0101013]}.
  
\bibitem{hep-th/0511254} 
  A.~Dymarsky, I.~R.~Klebanov and N.~Seiberg,
  ``On the moduli space of the cascading SU(M+p) x SU(p) gauge theory,''
  JHEP\ {\bf 0601}, 155  (2006)
  \href{http://arxiv.org/abs/hep-th/0511254}{[hep-th/0511254]}.
  
\bibitem{hep-th/0405282} 
  S.~S.~Gubser, C.~P.~Herzog and I.~R.~Klebanov,
  ``Symmetry breaking and axionic strings in the warped deformed conifold,''
  JHEP\ {\bf 0409}, 036  (2004)
  \href{http://arxiv.org/abs/hep-th/0405282}{[hep-th/0405282]}.
  
\bibitem{hep-th/0409186} 
  S.~S.~Gubser, C.~P.~Herzog and I.~R.~Klebanov,
  ``Variations on the warped deformed conifold,''
  Comptes Rendus Physique\ {\bf 5}, 1031  (2004)
  \href{http://arxiv.org/abs/hep-th/0409186}{[hep-th/0409186]}.
  
\bibitem{arXiv:0803.1315} 
  M.~K.~Benna and I.~R.~Klebanov,
  ``Gauge-String Dualities and Some Applications,''
  \href{http://arxiv.org/abs/0803.1315}{arXiv:0803.1315 [hep-th]}.
  
\bibitem{arXiv:1111.6567} 
  N.~Halmagyi, J.~T.~Liu and P.~Szepietowski,
  ``On N = 2 Truncations of IIB on $T^{1,1}$,''
  \href{http://arxiv.org/abs/1111.6567}{arXiv:1111.6567 [hep-th]}.
  
\bibitem{arXiv:1008.0983} 
  I.~Bena, G.~Giecold, M.~Grana, N.~Halmagyi and F.~Orsi,
  ``Supersymmetric Consistent Truncations of IIB on $T^{1,1}$,''
  JHEP\ {\bf 1104}, 021  (2011)
  \href{http://arxiv.org/abs/1008.0983}{[arXiv:1008.0983 [hep-th]]}.
  
\bibitem{hep-th/0206098} 
  V.~Borokhov and S.~S.~Gubser,
  ``Nonsupersymmetric deformations of the dual of a confining gauge theory,''
  JHEP\ {\bf 0305}, 034  (2003)
  \href{http://arxiv.org/abs/hep-th/0206098}{[hep-th/0206098]}.
  
\bibitem{hep-th/0309011} 
  S.~Kuperstein and J.~Sonnenschein,
  ``Analytic nonsupersymmtric background dual of a confining gauge theory and the corresponding plane wave theory of hadrons,''
  JHEP\ {\bf 0402}, 015  (2004)
  \href{http://arxiv.org/abs/hep-th/0309011}{[hep-th/0309011]}.
  
\bibitem{arXiv:0912.3519} 
  I.~Bena, M.~Grana and N.~Halmagyi,
  ``On the Existence of Meta-stable Vacua in Klebanov-Strassler,''
  JHEP\ {\bf 1009}, 087  (2010)
  \href{http://arxiv.org/abs/0912.3519}{[arXiv:0912.3519 [hep-th]]}.
  
\bibitem{arXiv:1106.6165} 
  I.~Bena, G.~Giecold, M.~Grana, N.~Halmagyi and S.~Massai,
  ``The backreaction of anti-D3 branes on the Klebanov-Strassler geometry,''
  \href{http://arxiv.org/abs/1106.6165}{arXiv:1106.6165 [hep-th]}.
  
\bibitem{arXiv:1102.2403} 
  I.~Bena, G.~Giecold, M.~Grana, N.~Halmagyi and S.~Massai,
  ``On Metastable Vacua and the Warped Deformed Conifold: Analytic Results,''
  \href{http://arxiv.org/abs/1102.2403}{arXiv:1102.2403 [hep-th]}.
  
\bibitem{arXiv:1102.1734} 
  A.~Dymarsky,
  ``On gravity dual of a metastable vacuum in Klebanov-Strassler theory,''
  JHEP\ {\bf 1105}, 053  (2011)
  \href{http://arxiv.org/abs/1102.1734}{[arXiv:1102.1734 [hep-th]]}.
  
\bibitem{arXiv:1011.2626} 
  I.~Bena, G.~Giecold, M.~Grana and N.~Halmagyi,
  ``On The Inflaton Potential From Antibranes in Warped Throats,''
  \href{http://arxiv.org/abs/1011.2626}{arXiv:1011.2626 [hep-th]}.
  
\bibitem{arXiv:1011.2195} 
  I.~Bena, G.~Giecold and N.~Halmagyi,
  ``The Backreaction of Anti-M2 Branes on a Warped Stenzel Space,''
  JHEP\ {\bf 1104}, 120  (2011)
  \href{http://arxiv.org/abs/1011.2195}{[arXiv:1011.2195 [hep-th]]}.
  
\bibitem{arXiv:1108.1789} 
  G.~Giecold, E.~Goi and F.~Orsi,
  ``Assessing a candidate IIA dual to metastable supersymmetry-breaking,''
  JHEP {\bf 1202}, 019 (2012)
  \href{http://arxiv.org/abs/1108.1789}{[arXiv:1108.1789 [hep-th]]}.
  
\bibitem{arXiv:1110.2513} 
  S.~Massai,
  ``Metastable Vacua and the Backreacted Stenzel Geometry,''
  \href{http://arxiv.org/abs/1110.2513}{arXiv:1110.2513 [hep-th]}.
  
\bibitem{arXiv:1106.0002} 
  S.~Gandhi, L.~McAllister and S.~Sjors,
  ``A Toolkit for Perturbing Flux Compactifications,''
  JHEP {\bf 1112}, 053 (2011)
  \href{http://arxiv.org/abs/1106.0002}{[arXiv:1106.0002 [hep-th]]}.
  
\bibitem{arXiv:1111.1727} 
  S.~Bennett, E.~Caceres, C.~Nunez, D.~Schofield and S.~Young,
  ``The Non-SUSY Baryonic Branch: Soft Supersymmetry Breaking of N=1 Gauge Theories,''
  \href{http://arxiv.org/abs/1111.1727}{arXiv:1111.1727 [hep-th]}.
  
\bibitem{arXiv:1111.1731} 
  A.~Dymarsky and S.~Kuperstein,
  ``Non-supersymmetric Conifold,''
  \href{http://arxiv.org/abs/1111.1731}{arXiv:1111.1731 [hep-th]}.
  
\bibitem{arXiv:1105.4879} 
  J.~Blaback, U.~H.~Danielsson, D.~Junghans, T.~Van Riet, T.~Wrase and M.~Zagermann,
  ``The problematic backreaction of SUSY-breaking branes,''
  JHEP\ {\bf 1108}, 105  (2011)
  \href{http://arxiv.org/abs/1105.4879}{[arXiv:1105.4879 [hep-th]]}.
  
\bibitem{arXiv:1111.2605} 
  J.~Blaback, U.~H.~Danielsson, D.~Junghans, T.~Van Riet, T.~Wrase and M.~Zagermann,
  ``(Anti-)Brane backreaction beyond perturbation theory,''
  JHEP {\bf 1202}, 025 (2012)
  \href{http://arxiv.org/abs/1111.2605}{[arXiv:1111.2605 [hep-th]]}.
  
\bibitem{HoyosBadajoz:2008fw} 
  C.~Hoyos-Badajoz, C.~Nunez and I.~Papadimitriou,
  ``Comments on the String dual to N=1 SQCD,''
  Phys.\ Rev.\ D {\bf 78}, 086005 (2008)
  \href{http://arxiv.org/abs/0807.3039}{[arXiv:0807.3039 [hep-th]]}.
  
\bibitem{arXiv:0906.0591} 
  J.~Maldacena and D.~Martelli,
  ``The Unwarped, resolved, deformed conifold: Fivebranes and the baryonic branch of the Klebanov-Strassler theory,''
  JHEP\ {\bf 1001}, 104  (2010)
  \href{http://arxiv.org/abs/0906.0591}{[arXiv:0906.0591 [hep-th]]}.
  
\bibitem{hep-th/0108101} 
  C.~P.~Herzog, I.~R.~Klebanov and P.~Ouyang,
  ``Remarks on the warped deformed conifold,''
  \href{http://arxiv.org/abs/hep-th/0108101}{hep-th/0108101}.
  
\bibitem{hep-th/0205100} 
  C.~P.~Herzog, I.~R.~Klebanov and P.~Ouyang,
  ``D-branes on the conifold and N=1 gauge / gravity dualities,''
  \href{http://arxiv.org/abs/hep-th/0205100}{hep-th/0205100}.
  
\bibitem{PDEs}
Y.~Pinchover and J.~Rubinstein,
``\href{http://books.google.com/books?id=CnvDS9twvUMC&printsec=frontcover#v=onepage&q&f=false}{An Introduction to Partial Differential Equations},''
  C.U.P. (2005)
  
\bibitem{PDEs2}
L.~C.~Evans,
``\href{http://books.google.com/books?id=Xnu0o_EJrCQC&printsec=frontcover#v=onepage&q&f=false}{Partial Differential Equations},''
  A.M.S. (2010)
  
\bibitem{unpub}
 G.~Giecold and N.~Halmagyi, unpublished (2011).  
 
\bibitem{Ceresole:1999zg} 
  A.~Ceresole, G.~Dall'Agata, R.~D'Auria and S.~Ferrara,
  ``M theory on the Stiefel manifold and 3-D conformal field theories,''
  JHEP {\bf 0003}, 011 (2000)
  \href{http://arxiv.org/abs/hep-th/9912107}{[hep-th/9912107]}.
  
\bibitem{Martelli:2009ga} 
  D.~Martelli and J.~Sparks,
  ``AdS(40 / CFT(3) duals from M2-branes at hypersurface singularities and their deformations,''
  JHEP {\bf 0912}, 017 (2009)
  \href{http://arxiv.org/abs/0909.2036}{[arXiv:0909.2036 [hep-th]]}.
  
\bibitem{Klebanov:2010qs} 
  I.~R.~Klebanov and S.~S.~Pufu,
  ``M-Branes and Metastable States,''
  JHEP {\bf 1108}, 035 (2011)
  \href{http://arxiv.org/abs/1006.3587}{[arXiv:1006.3587 [hep-th]]}.
  
\bibitem{Chemissany:2010zp} 
  W.~Chemissany, P.~Fre, J.~Rosseel, A.~S.~Sorin, M.~Trigiante and T.~Van Riet,
  ``Black holes in supergravity and integrability,''
  JHEP {\bf 1009}, 080 (2010)
  \href{http://arxiv.org/abs/1007.3209}{[arXiv:1007.3209 [hep-th]]}.

\end{thebibliography}
\end{document}